\documentclass{v18windows}

\usepackage{amsmath}

\def\be{\begin{equation}}
\def\ee{\end{equation}}
\def\bea{\begin{eqnarray}}
\def\eea{\end{eqnarray}}

\newcommand{\eq}[1]{Eq.~(\ref{#1})}
\newcommand{\bib}[1]{Ref.~\cite{#1}}

\newcommand{\fig}[1]{Fig.~\ref{#1}}
\newcommand{\tab}[1]{Table~\ref{#1}}

\newcommand{\crn}{\nonumber \\}

\newcommand{\fr}{\frac}

\newcommand{\Photo}{}

\begin{document}
\rightline{IFIRSE-TH-2018-3}
\vspace*{4cm}
\title{{\boldmath $WZ$} PRODUCTION AT THE LHC: POLARIZATION OBSERVABLES IN THE STANDARD MODEL}

\author{ JULIEN BAGLIO $^{1,2}$, LE DUC NINH $^{2}$ \footnote{Speaker} }

\address{$^1$ Institut f\"{u}r Theoretische Physik, Eberhard Karls Universit\"{a}t
T\"{u}bingen,\\ Auf der Morgenstelle 14, D-72076 T\"{u}bingen, Germany}
\address{$^2$ Institute for Interdisciplinary Research in Science and Education,\\  
ICISE, 590000 Quy Nhon, Vietnam}

\maketitle\abstracts{
$WZ$ production is an important process at the LHC 
because it probes the non-Abelian structure
of electroweak interactions and it is a background process for many new physics 
searches. In the quest for new physics, 
polarization observables of the $W$ and $Z$ 
bosons can play an important role. 
They can be extracted from measurements 
and can be calculated using the Standard Model. 
In this contribution, we define fiducial polarization observables of the 
$W$ and $Z$ bosons and discuss the effects of next-to-leading order 
electroweak and QCD corrections in the Standard Model.}

\section{Introduction}

The process $pp \to W^\pm Z \to 3\ell + \nu$ is important in the physics
program at the Large Hadron Collider (LHC). It has been extensively
studied by both theorists and experimentalists. At leading order (LO), it is 
sensitive to the triple-gauge couplings. With high statistics, 
precision measurements can be performed, thereby allowing precise 
comparisons between theoretical predictions and measurements 
for non-trivial observables such as kinematical distributions and 
polarization observables of the massive gauge bosons. This 
is important to find new physics effects. 

At the LHC, the initial proton beams are unpolarized. 
However, the $W$ or $Z$ bosons produced there are intrinsically 
polarized because of the asymmetry in their interaction with left- and right-handed 
quarks. The $W$ bosons couple only to left-handed quarks, while  
the $Z$ boson interacts with both left- and right-handed quarks, but with different
coupling strengths. 

In the following we will first discuss how polarization observables are defined from the 
spin-density matrix of a massive gauge boson. This is established at LO and without 
kinematical cut. However, in order to have precise comparisons with measurements, higher-order 
effects and kinematical cuts on the decay leptons must be taken into account. 
This leads us to the definition of fiducial polarization observables, that will 
be shown to have similar characteristics 
to the true polarization observables and can be easily calculated from the polar-azimuthal 
angle distribution of the decay lepton using projections. No template fitting is needed. 
The angular distribution is calculated as usual using fiducial phase-space cuts.  

\section{Fiducial polarization observables}
\label{sec:fiducial_polarization}
A polarized on-shell massive gauge boson can be described by a $3\times 3$ spin-density matrix $\rho_{ij}$. 
This matrix is Hermitian and satisfies the normalization condition of $\text{Tr}(\rho)=1$, 
hence it can be parameterized by $8$ real parameters. 

We consider the decay of a {\em polarized} massive gauge boson to two leptons $V\to l_1 l_2$. 
For the case of the $W$ bosons we have $W^\pm \to e^\pm \nu_e$ and for the $Z$ boson 
$Z\to \mu^- \mu^+$. The amplitude squared reads
\begin{align}
|\mathcal{M}|^2 = 
\sum_{i,j=1}^{3}\sum_{\lambda_1,\lambda_2=1}^{2}\rho_{ij}\mathcal{M}_{i\lambda_1\lambda_2}\mathcal{M}^*_{j\lambda_1\lambda_2}.
\label{eq:polarized_decay_V}
\end{align}
This equation is also correct for off-shell $V$ if the lepton masses 
are neglected. An off-shell $V$ can be described by 4 polarization vectors, two transverse, 
one longitudinal and one auxiliary mode. However, the auxiliary vector can be chosen 
to be proportional the the momentum of the gauge boson, 
hence its contribution vanishes in the limit of massless leptons. 

In order to link the spin-density matrix to the angular distribution of a decay lepton, say $l_1$, 
we have to know the dependence of the helicity amplitude $\mathcal{M}_{i\lambda_1\lambda_2}$ on 
the angles $\phi_1$ and $\theta_1$. At LO, we have \cite{Aguilar-Saavedra:2015yza,Aguilar-Saavedra:2017zkn}
\begin{align}
\mathcal{M}_{m\lambda_1\lambda_2} &= a_{\lambda_1\lambda_2} e^{im\phi} d^{1}_{m\lambda}(\theta), \; \lambda = \lambda_1 - \lambda_2 = \pm 1,\\
d^1_{11}(\theta) &= \fr{1+\cos\theta}{2},\; d^1_{1-1}(\theta) = \fr{1-\cos\theta}{2},\;
d^1_{01}(\theta) = \fr{\sin\theta}{\sqrt{2}},\; d^j_{m'm} = (-1)^{m-m'}d^j_{mm'} = d^j_{-m-m'},\nonumber
\end{align}
where $a_{\lambda_1\lambda_2}$ are constants, $\phi$ and $\theta$ are the azimuthal and polar angles of 
the $l_1$ momentum in the $V$ boson rest frame. We then get the following angular distribution
\begin{align}
\fr{d\Gamma}{\Gamma d\!\cos\theta d\phi} &= \fr{3}{16\pi}
\Big[ 
(1+\cos^2\theta) + \hat{A}_0 \fr{1}{2}(1-3\cos^2\theta)
+ \hat{A}_1 \sin(2\theta)\cos\phi  \crn
& + \hat{A}_2 \fr{1}{2} \sin^2\theta \cos(2\phi)
+ \hat{A}_3 \sin\theta\cos\phi + \hat{A}_4 \cos\theta \crn
& + \hat{A}_5 \sin^2\theta \sin(2\phi) 
+ \hat{A}_6 \sin(2\theta) \sin\phi + \hat{A}_7 \sin\theta \sin\phi
\Big],\label{eq:definition_Ai_tot}
\end{align}
where $\hat{A}_{i}$ are dimensionless angular coefficients related to the spin-density matrix as
\begin{align}
\hat{A}_0 &= 2 \rho_{00},\; \hat{A}_1 = \fr{1}{\sqrt{2}}(\rho_{+0}-\rho_{-0}+\rho_{0+}-\rho_{0-}),\crn
\hat{A}_2 &= 2(\rho_{+-} + \rho_{-+}),\; \hat{A}_3 = \sqrt{2}b(\rho_{+0} + \rho_{-0} + \rho_{0+} + \rho_{0-}),\crn
\hat{A}_4 &= 2b(\rho_{++} - \rho_{--}),\; \hat{A}_5 = \fr{1}{i}(\rho_{-+} - \rho_{+-}),\crn
\hat{A}_6 &= -\fr{1}{i\sqrt{2}}(\rho_{+0} + \rho_{-0} - \rho_{0+} - \rho_{0-}),\; 
\hat{A}_7 = \fr{\sqrt{2}b}{i}(\rho_{0+} - \rho_{0-} - \rho_{+0} + \rho_{-0}),
\label{eq:relations_Ai_spin_matrix}
\end{align}  
where $b=1$ for the $W^\pm$ bosons and $b=-c$ for the $Z$ boson, with
\bea
c = \fr{g_L^2 - g_R^2}{g_L^2 + g_R^2} =
\fr{1-4s^2_W}{1-4s^2_W+8s^4_W},\quad s^2_W = 1 - \fr{M_W^2}{M_Z^2}.
\label{eq:def_c}
\eea
Another way to calculate the coefficients $\hat{A}_{i}$ is using various angular projections of the distribution 
$d\Gamma/(\Gamma d\!\cos\theta d\phi)$ as \cite{Bern:2011ie,Stirling:2012zt}
\begin{align}
\hat{A}_0 &= 4 - \langle 10\cos^2\theta\rangle, \; \hat{A}_1 = \langle 5\sin 2\theta\cos\phi \rangle, 
\; \hat{A}_2 = \langle 10\sin^2\theta \cos 2\phi \rangle, \; \hat{A}_3 = \langle 4 \sin\theta\cos\phi \rangle,\crn 
\; \hat{A}_4 &= \langle 4\cos\theta\rangle, \; \hat{A}_5 = \langle 5\sin^2\theta\sin 2\phi \rangle, 
\; \hat{A}_6 = \langle 5\sin 2\theta \sin\phi \rangle, \; \hat{A}_7 = \langle 4\sin\theta \sin\phi \rangle,
\label{eq:cal_Ai}
\end{align}
with the following definition for angular projection
\begin{align}
\langle g(\theta,\phi) \rangle = \int_{-1}^{1} d\!\cos\theta \int_{0}^{2\pi} d\!\phi
g(\theta,\phi)\fr{d\Gamma}{\Gamma d\!\cos\theta d\phi}.
\label{eq:defs_expectation_3D} 
\end{align}
This second way of calculating the polarization observables is very powerful as it allows us to define 
fiducial polarization observables by replacing $d\Gamma/(\Gamma d\!\cos\theta d\phi)$ in \eq{eq:defs_expectation_3D} 
by the corresponding fiducial distribution $d\sigma/(\sigma d\!\cos\theta d\phi)$ \cite{Stirling:2012zt}. 
The new angular coefficients are 
now denoted $A_i$, i.e. without the hat. We call them fiducial polarization observables or angular coefficients. 
The key differences between 
$d\Gamma/(\Gamma d\!\cos\theta d\phi)$ and $d\sigma/(\sigma d\!\cos\theta d\phi)$ are that the former includes 
only Feynman diagrams with $W\to \ell \nu_\ell$ or $Z\to l^- l^+$ and no cut on the individual leptons is allowed, 
while those restrictions are lifted for the latter. $d\sigma/(\sigma d\!\cos\theta d\phi)$ is a normal 
angular distribution, which is calculated using 
the full matrix elements including off-shell, interference and higher-order effects and with arbitrary cuts on the individual leptons. 

The fiducial polarization fractions are defined as
\begin{align}
f^V_L 
  = -\fr{1}{2} + d\langle\cos\theta\rangle +
    \fr{5}{2}\langle\cos^2\theta\rangle, \;
f^V_R 
  = -\fr{1}{2} - d\langle\cos\theta\rangle +
    \fr{5}{2}\langle\cos^2\theta\rangle, \;
f^V_0 
  = 2 - 5 \langle\cos^2\theta\rangle,
\label{eq:cal_fLR0_V}
\end{align}
where $d=\mp 1 $, $\theta = \theta_{e}$ for $V=W^\pm$ and $d=1/c$, $\theta = \theta_{\mu^-}$ for $V=Z$. 
\section{Numerical results}
\label{sec:results}
In this work, the fiducial distribution $d\sigma/(\sigma d\!\cos\theta d\phi)$ 
is calculated at LO and next-to-leading order (NLO) QCD where the full matrix elements 
are used. The NLO QCD results are obtained using the {\tt VBFNLO} program \cite{Arnold:2008rz}. 
The NLO electroweak (EW) corrections are calculated using the double pole approximation where EW 
corrections to the on-shell $WZ$ production and to the decays $W\to e \nu_e$, $Z\to \mu^+ \mu^-$ 
are separately included. The EW corrections to the on-shell production part are taken over from \bib{Baglio:2013toa}. 
Further calculation details are provided in \bib{Baglio:2018rcu}.
\begin{table}[hb!]
  \renewcommand{\arraystretch}{1.3}
\begin{center}
    \fontsize{8}{8}
\begin{tabular}{|c|c|c|c||c|c|c|}\hline
$\text{Method}$  & $f^{W^+}_L$ & $f^{W^+}_0$ & $f^{W^+}_R$ & $f^Z_L$ & $f^Z_0$ & $f^Z_R$\\
\hline
$\text{HE NLOQCD}$ & $0.320(2)^{+2}_{-2}$ & $0.508(1)^{+2}_{-2}$ & $0.172(2)^{+4}_{-3}$ & $0.257(1)^{+3}_{-3}$ & $0.493(1)^{+2}_{-3}$ & $0.251(1)^{+1}_{-0.5}$\\
\hline
$\text{HE NLOQCDEW}$ & $0.320$ & $0.507$ & $0.173$ & $0.254$ & $0.493$ & $0.253$\\
\hline
\end{tabular}
    \caption{\small $W^+_{}$ and $Z$ fiducial polarization fractions at NLO QCD and
  NLO QCD+EW in the helicity (HE) 
  coordinate system. The PDF uncertainties (in parenthesis) and the scale
  uncertainties are provided for the NLO QCD results.}
    \label{tab:coeff_fL0R_WpZ_ATLAS}
\end{center}
\end{table}
\begin{table}[ht!]
 \renewcommand{\arraystretch}{1.3}
\begin{center}
\setlength\tabcolsep{0.03cm}
\fontsize{8.0}{8.0}
\begin{tabular}{|c|c|c|c|c|c|c|c|c|}\hline
$\text{Method}$  & $A_0$ & $A_1$  & $A_2$ & $A_3$ & $A_4$ & $A_5$ & $A_6$ & $A_7$\\
\hline
{\fontsize{6.0}{6.0}$\text{HE NLOQCD}$} & $0.985(2)^{+5}_{-6}$ & $-0.306(1)^{+4}_{-3}$ & $-0.734(1)^{+2}_{-2}$ & $0.031(1)^{+2}_{-2}$ & $0.003(1)^{+1}_{-1}$ & $-0.004(1)^{+0.3}_{-0.4}$ & $-0.004(1)^{+0.3}_{-0.2}$ & $0.003(1)^{+0.2}_{-0}$\\
\hline
{\fontsize{6.0}{6.0}$\text{HE NLOQCDEW}$} & $0.986$ & $-0.306$ & $-0.738$ & $0.024$ & $0.0001$ & $-0.004$ & $-0.004$ & $0.003$\\
\hline
\end{tabular}
\caption{\small Same as \tab{tab:coeff_fL0R_WpZ_ATLAS} but for the $Z$ fiducial angular coefficients.}
\label{tab:coeff_Ai_Z_ATLAS}
\end{center}
\end{table}

We present here results for fiducial polarization observables for the process 
$pp \to e^+ \nu_e\, \mu^+ \mu^- + X$ at the 13 TeV LHC with the ATLAS fiducial cuts 
defined in \bib{Aaboud:2016yus}. Fiducial polarization fractions of the $W^+$ and $Z$ bosons 
are presented in \tab{tab:coeff_fL0R_WpZ_ATLAS}, while angular coefficients of the $Z$ boson 
are in \tab{tab:coeff_Ai_Z_ATLAS}, in the helicity coordinate system \cite{Bern:2011ie}. 
We observe that the coefficients $A_5$, $A_6$ and $A_7$ are very small. This is understandable 
as they are proportional to the imaginary parts of the spin-density matrix in the on-shell approximation, 
see \eq{eq:relations_Ai_spin_matrix}. The NLO EW corrections are negligible for $A_0$, $A_1$ and $A_2$, 
but are significant for $A_3$ and $A_4$. For the case of the $W^\pm$ bosons, EW corrections are negligible \cite{Baglio:2018rcu}. 
It is interesting to note that $A^Z_3$ and $A^Z_4$ are proportional to the parameter $c$ which is a function of $s^2_W$, 
see \eq{eq:relations_Ai_spin_matrix}. The origin of large EW corrections to $A^Z_3$ and $A^Z_4$ is traced 
back to the radiative corrections to the $Z\to \mu^- \mu^+$ decay.

\begin{figure}[ht!]
  \centering
  \begin{tabular}{cc}
    \includegraphics[width=0.48\textwidth]{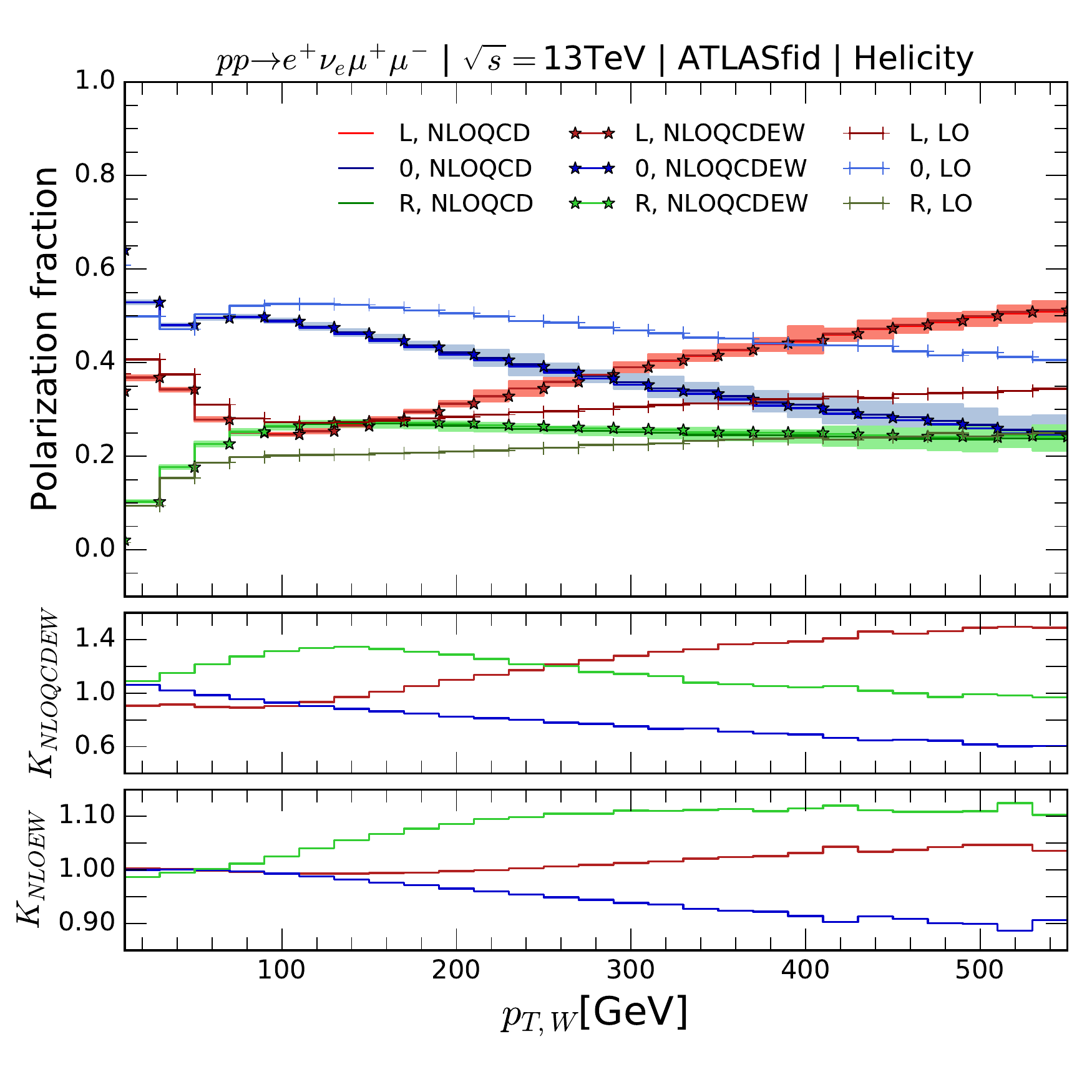}& 
    \includegraphics[width=0.48\textwidth]{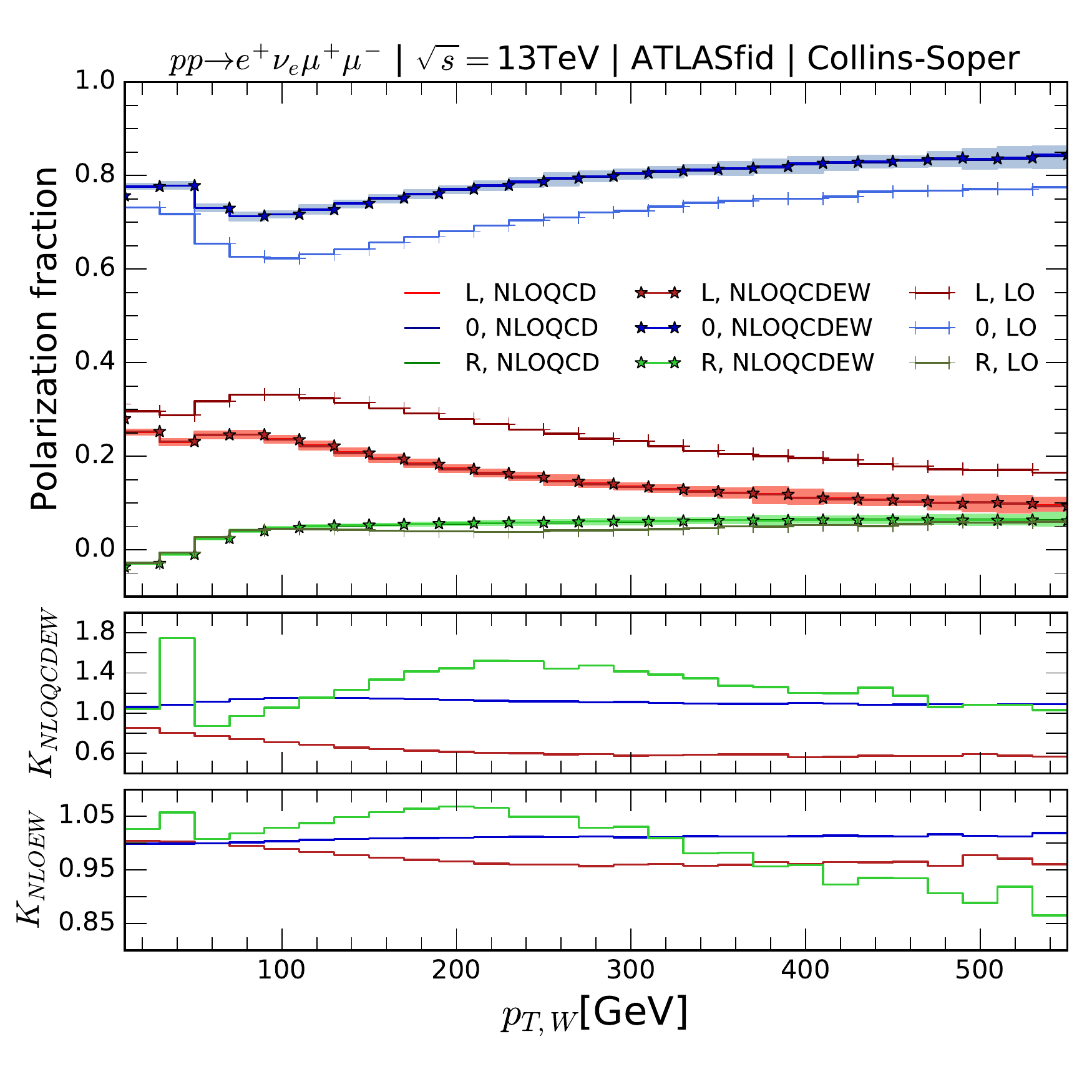}
  \end{tabular}
  \caption{Transverse momentum distributions of the $W^+$ boson fiducial polarization fractions. 
    The left-hand-side plot is for the helicity coordinate system, while the right-hand-side
    plot is for the Collins-Soper coordinate system. The bands include PDF and
    scale uncertainties calculated at NLOQCD using linear summation. The $K$ factors in the small panels are 
    the ratios of the NLO result over the LO one.}
  \label{fig:dist_ptWp_ATLAS_NLOQCDEW}
\end{figure}

The dependence of the fiducial polarization fractions on $p_{T,W}$ is shown in \fig{fig:dist_ptWp_ATLAS_NLOQCDEW} 
for the helicity (left) and Collins-Soper (right) coordinate systems. For the Collins-Soper system, 
the $f_R$ fraction gets negative at low $p_{T,W}$ and the longitudinal fraction does not decrease at high energies. 
For the helicity system, the fractions are positive and the longitudinal fraction decreases with high $p_{T,W}$ 
as suggested by the equivalence theorem. Similar behavior is seen in the $p_{T,Z}$ dependence of the $Z$ polarization fractions. 
We therefore conclude that the helicity coordinate system is the better choice for 
calculating (fiducial) polarization observables for both the $W$ and $Z$ bosons. 

\section*{Acknowledgments}
JB acknowledges the support from the
Carl-Zeiss foundation. This research is funded by the Vietnam
National Foundation for Science and Technology Development (NAFOSTED)
under grant number 103.01-2017.78. LDN thanks the organizers for the invitation.


\end{document}